# Distributed SDN-Based Communication Architecture for the Pods4Rail System

Dingyang Liu[1][0000-0002-8988-6543], Dereje Mechal Molla[1*][0000-0003-4515-679X], Léo Mendiboure[1,2][0000-0001-6643-9567], and Marion Berbineau[1][0000-0003-3807-9669]

[1] Université Gustave Eiffel, COSYS-LEOST, Villeneuve d'Ascq, France
*dereje.molla2@ univ-eiffel.fr,
marion.berbineau@univ-eiffel.fr
[2] Université de Pau et des Pays de l'Adour, 64000 Pau, France,
leo.mendiboure@univ-pau.fr

**Abstract.** Future multimodal transportation systems require reliable, low-latency communication infrastructures to coordinate autonomous vehicles and moving infrastructure across rail and road networks. Traditional centralized control architectures struggle to meet these requirements in highly dynamic environments due to increased latency, limited scalability, and poor adaptability to changing network conditions. To address this, we propose a distributed communication architecture integrating Software-Defined Networking (SDN) and Multi-Access Edge Computing (MEC), that can create a flexible, programmable and low-latency network. Results show controller communication and flow setup latency between edge SDN controllers and Pods are lower than reported in literature. The framework uses hierarchical control with regional and edge controllers to support low-latency interface management and edge autonomy. Operational workflows and control logic are defined as representative scenarios. The architecture combines regional policy coordination with edge-level autonomy, enabling local failover and adaptive interface management without central dependency.

## 1 Introduction

Intelligent Transportation Systems (ITS) are moving towards seamless multimodal integration to improve efficiency, safety, and sustainability. A key challenge lies in enabling coordinated mobility services across heterogeneous infrastructures including rail, road, and ropeway networks. Recent French initiatives like Flexy [1] and FlexMove [2], as well as European efforts, highlight the potential of intermodal solutions to address last-mile challenges. At the European level, the Pods4Rail project [3] introduces Pods (Pod is a vehicle consisting of a carrier and transport unit) as moving vehicles, primarily for rail, while also supporting road and ropeway.

Automated and interoperable systems impose stringent communication requirements, especially for cross-scenario and cross-node coordination. Pods exchange information not only with each other but also with passengers and roadside/trackside units (RSU/TSU). This requires low-latency, programmable, and scalable communication that integrates with existing vehicular and railway networks. These requirements can be addressed by combining Software-Defined Networking (SDN), Network Functions Virtualization (NFV), Network Slicing (NS), and Multi-Access Edge Computing (MEC) [4].

Conventional centralized SDN architectures face critical limitations in highly dynamic wireless and transportation environments. Central controllers may become bottlenecks if they introduce single points of failure, and respond too slowly to sudden changes [5]. These shortcomings hinder their ability to support the fast, resilient, and distributed coordination required for Pods mobility. To address these issues, this paper proposes a distributed SDN-enabled communication architecture tailored to the Pods4Rail system. The framework leverages MEC to shift part of the control intelligence from centralized entities to edge nodes (*e.g.*, RSU/TSU), enabling local decision-making, flexible radio interface (RAT) management, and real-time coordination [6]. The design ensures continued operation even under link failures or rapid topology changes by delegating decision-making to edge nodes. In summary, this work introduces a conceptual architecture for Pods4Rail that supports scheduling, collaboration, and transport layers, ensuring modular and adaptive communication; presents a distributed SDN framework that integrates edge controllers with regional coordination; and proposes two control mechanisms for critical scenarios: dynamic virtual coupling between Pods and adaptive RAT selection during link failures.

The remainder of this paper is structured as follows: Section 2 presents the Pods4Rail system architecture and communication workflow. Section 3 reviews related work. Section 4 details the distributed SDN framework and control mechanisms. Section 5 concludes with perspectives for future research.

## 2 Architecture of the Pods4Rail System

Pods4Rail is a forward-looking transportation project designed to deliver flexible, automated, and sustainable mobility. It promotes rail–road interoperability, combining the high capacity of rail with the flexibility of roads to meet diverse mobility demands, from urban commuting to logistics in remote areas. By extending rail services closer to departure and arrival points, Pods4Rail directly addresses the last-mile challenge [3]. The system aims to create a user-centric, efficient, and low-carbon mobility ecosystem. Its architecture is shown in Fig. 1 depicting its main capabilities include intelligent scheduling, collaborative autonomy, and flexible data transmission.

- Scheduling layer: the Dispatch Center (DC) manages transportation services through NS, which partitions physical resources into isolated virtual networks tailored to specific requirements such as latency, reliability, or bandwidth [7]. Using Base Stations or Edge Nodes (ECNs), the DC collects user requests (destination, estimated time, *etc*.,) and provides matching Pod information (location, arrival time, etc.,) to allocate tasks and paths.

- Collaboration layer: MEC platform aggregates local environment data from RSUs, cameras, and traffic lights. The embedded SDN controller processes perception data, distributes flow rules, schedules interfaces, and sorts tasks by priority. Existing studies, such as [8], proposed lightweight SDN controllers for dynamic RAT selection. For Pods, controllers must also interpret road behaviors (*e.g.*, virtual coupling) and guide Pods toward autonomous responses and regional coordination via localized command issuance.
- Transport layer: Pods achieve dynamic behaviors such as virtual coupling, decoupling, and reorganization through vehicle-to-vehicle (V2V) communication, improving road and rail traffic efficiency. For external connectivity, Pods use Vehicle-to-Network (V2N) links to access operator networks (*e.g.*, Operator 1/2), enabling cross-regional synchronization, remote command reception, and access control. In transfer scenarios, the system supports multimodal mechanisms, ensuring seamless interconnection with railways, buses, and other transport modes for end-to-end travel continuity.

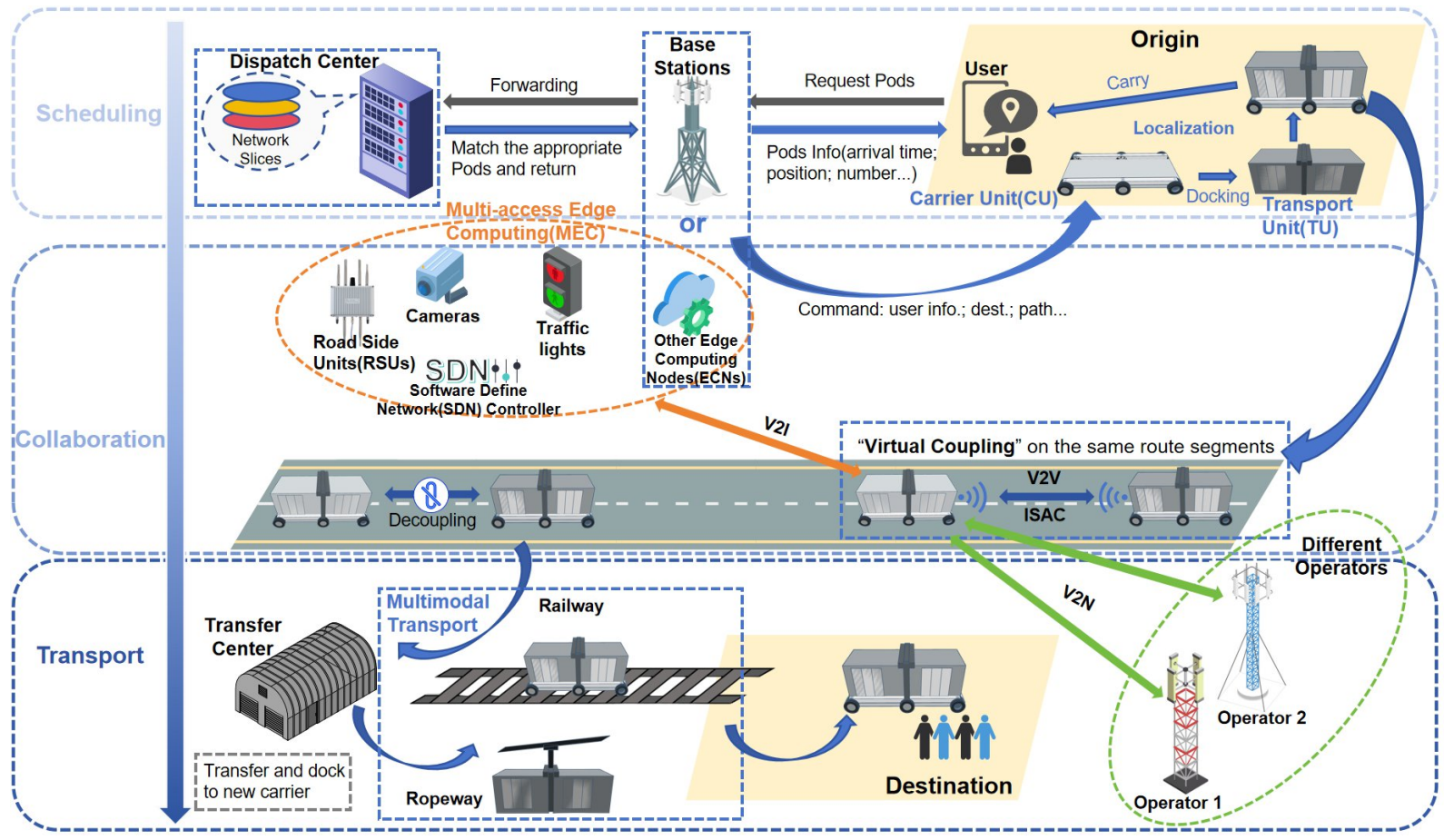

**Fig. 1.** Conceptual architecture for the Pods4Rail system.

## 3 Related Work

Several studies have explored SDN-based architectures for vehicular and intelligent transportation systems. In [9], the authors constructed a multi-level SDN control system comprising a global controller and local controllers, introducing an "Interface Selector" module for RAT selection. However, the focus of the architecture remains on the global controller, and its control granularity is concentrated on RAT selection and QoS decision-making. It does not utilize MEC to distribute computing resources.

Oubbati *et al*. [10] proposed and analyzed an SDN-based ITS to improve communication efficiency and path selection capabilities in VANETs, but most of the control commands are initiated by the central controller. Adbeb *et al*. [11] proposed an SD-VANET architecture that communicates with vehicles and RSUs via the OpenFlow protocol - the first SDN protocol proposed for the southbound interface (SBI) [12], collect traffic status information, and calculate optimal paths. Their work is mainly based on static flow table deployment and still centers on centralized SDN controllers.

These studies share two major limitations: first, most of the studies consider the traditional centralized control architecture based on the OpenFlow protocol. However, in a highly-dynamic scenarios, its limitations have become increasingly apparent: centralized dependency increases control plane pressure, static flow table matching lacks flexibility, and it struggles to adapt to rapid changes in complex network states. Second, although some efforts have been made to introduce a hierarchical mechanism, they have a limited scope of control. Their focus is restricted to communication interfaces and QoS, without extending to behavior scheduling, local recovery, or edge autonomy. In contrast, our framework addresses these gaps by deploying lightweight SDN controllers at edge nodes. By integrating these controllers with MEC for local processing, the framework enables Pods to manage interfaces, schedule behaviors, and recover from failures without immediate central input. Furthermore, while prior work focuses on optimizing central decisions or RAT selection, our approach introduces a distributed framework combining multi-interface management, behavior scheduling, and emergency recovery. This makes it particularly well-suited for Pods4Rail, where dynamic multimodal environments demand both global coordination and localized autonomy.

## 4 Distributed SDN Framework

Building on the overall communication design, this section introduces a distributed SDN framework to address the scalability and reliability limitations of centralized solutions in large-scale, dynamic environments. The framework is described at two levels: system-wide coordination through regional controllers, and local execution through edge controllers. Finally, two control mechanisms are presented to illustrate operation in critical scenarios.

### 4.1 System-Level SDN-Enabled Distributed Architecture

The Pods4Rail infrastructure, and thus its communication architecture, spans a large area with many distributed Pods and edge nodes. Relying on a single centralized controller would quickly lead to scalability bottlenecks, congestion, and single points of failure [13]. The system therefore adopts a hierarchical distributed design (Fig. 2). Unlike centralized SDN architectures, this design explicitly distributes decision-making to regional and edge controllers, thereby enhancing scalability and resilience.

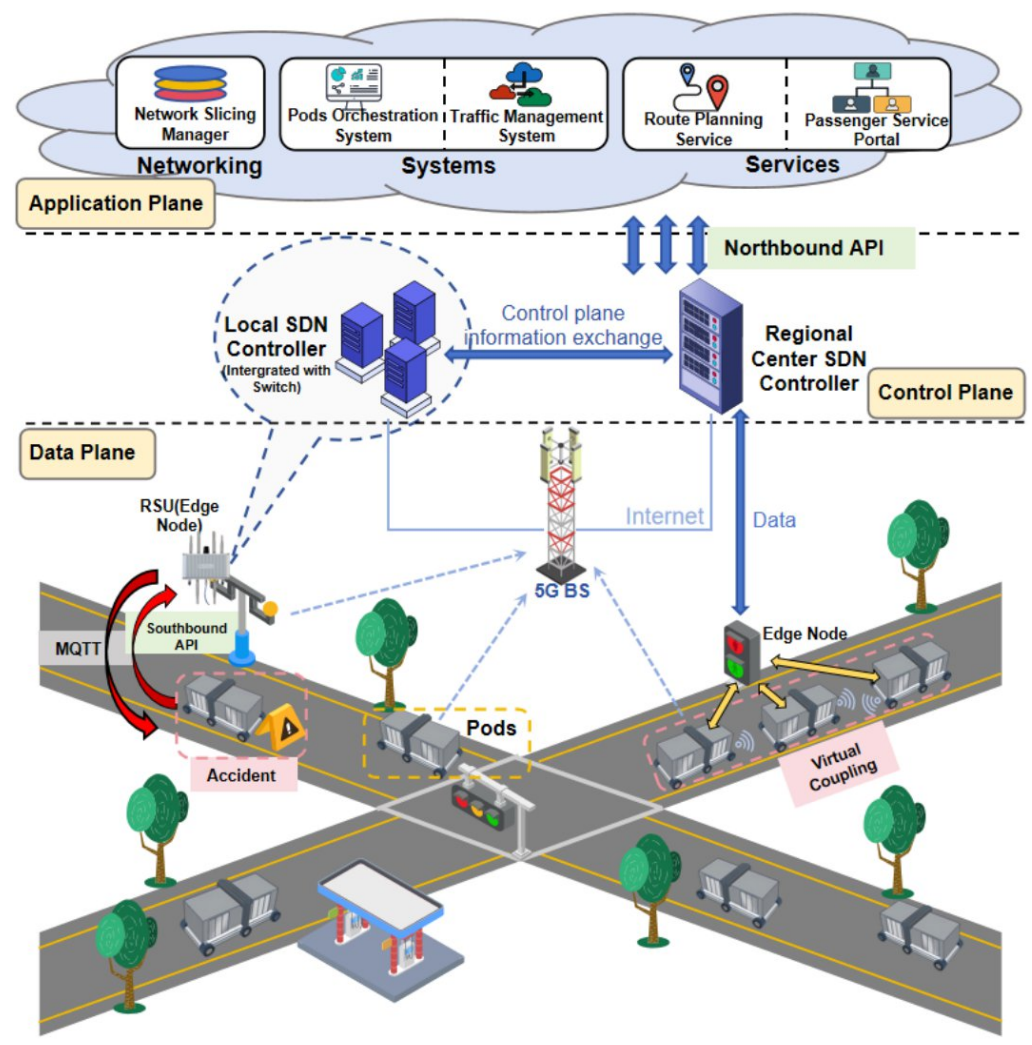

**Fig. 2.** Conceptual architecture for the Pods4Rail system.

The application plane defines high-level logic such as traffic management and route planning. The control plane is split between regional controllers, which coordinate policies and scheduling, and local controllers at edge communication nodes, which enforce policies and manage interfaces. The data plane comprises Pods, passenger devices, and other field units.

Local controllers interact with data-plane nodes via SBI. Since no standard protocol exists for mobile SDN in wireless environments, our previous work [8] proposed an MQTT-based lightweight controller, characterized by a reduced control logic footprint and event-driven SBI, that provides event notification, feedback, and collaborative scheduling.

### 4.2 Edge Node Execution Framework

While regional controllers coordinate policies, edge controllers ensure real-time local adaptation. Figure 3 illustrates the internal structure of ECNs in the Pods system. This architecture is centered on a lightweight SDN controller that integrates perception, decision-making, and execution. It provides unified management of multiple communication interfaces and behavior scheduling, with control logic interacting through the SBI and supported by local storage for policy caching, state recording, and rescheduling. When the regional controller is accessible, Pod nodes follow its unified scheduling for interface activation and forwarding. In case of controller failure, link disruption, or sudden events, nodes switch to local autonomy: the lightweight controller and interface switching module reconstruct paths and select optimal links. Status updates are then reported back through a reserved channel, ensuring rapid local response before central validation. Table 1 shows the time taken to communicate these data between edge SDN controller and Pods, obtained with the setup described in [8].

It shows that controller and flow setup latency are lower than reported in literature [14]. This structure ensures that Pods maintain essential connectivity and decision-making capacity even if higher-level control becomes temporarily unavailable.

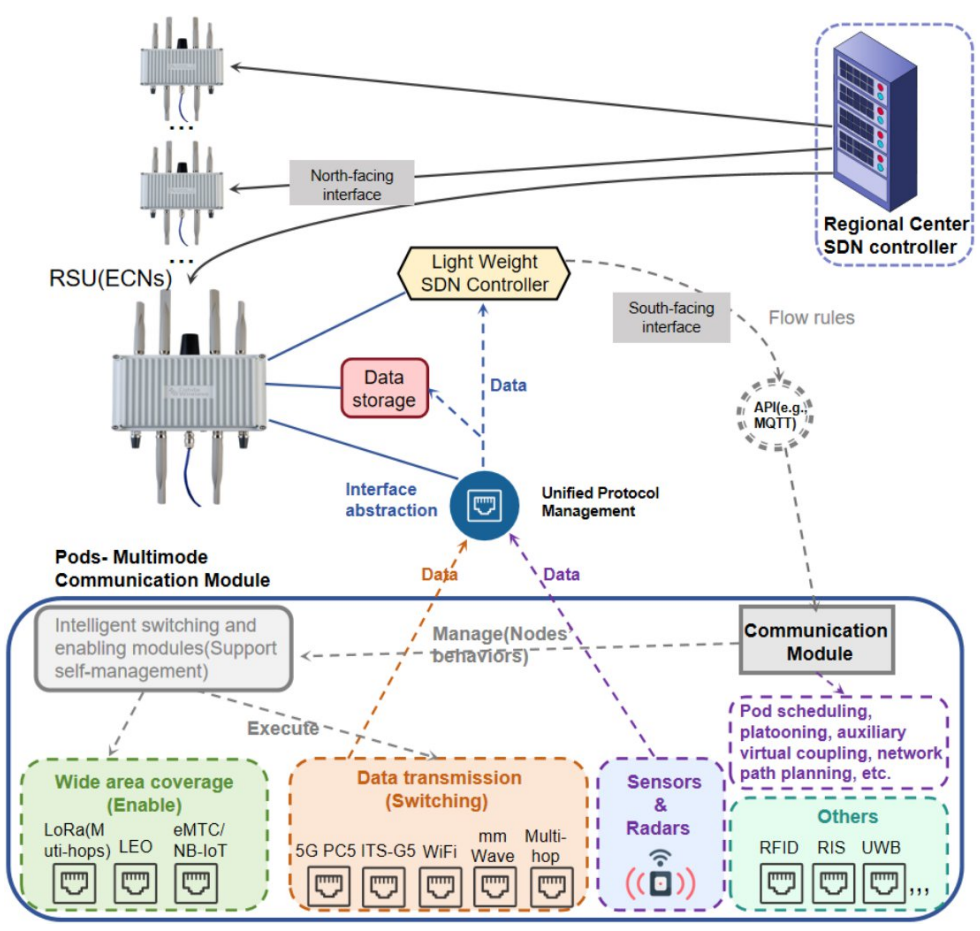

**Fig. 2.** Conceptual architecture for the Pods4Rail system.

**Table 1.** SDN controller and flow rule latency.

| Metrics | Min (ms) | Max (ms) | Mean (ms) | Median (ms) | Stdev (ms) |
|---|---|---|---|---|---|
| $T_{Controller}$ | 0.262 | 0.506 | 0.327 | 0.318 | 0.055 |
| $T_{FlowRule}$ | 21.823 | 32.912 | 23.637 | 23.456 | 2.163 |
| $T_{Total}$ | 22.085 | 33.418 | 23.964 | 23.774 | 2.218 |

The communication modules are divided into four categories: Wide-area coverage for status reporting and scheduling, relying on long-range, low-data-rate links such as LoRa, eMTC/NB-IoT, and LEO satellites; Data transmission including 5G PC5, ITS-G5, Wi-Fi, mmWave, and multi-hop V2X; Environmental perception to support virtual coupling and fleet synchronization via real-time sensing; and Auxiliary modules for localization, path planning, and security such as RFID, RIS, and UWB.

### 4.3 Coordination in Critical Scenarios

After defining the distributed structure and functional roles of SDN at both the system and node levels, this section introduces two control mechanisms tailored to typical Pods operation scenarios, with the aim of enhancing coordination and communication continuity in dynamic environments.

**Virtual Coupling.** The scenario focuses on the formation and management of virtual coupling between Pods. When two Pods are detected to be traveling along overlapping route segments within the same time window, the edge controller constructs a candidate pair and reports it to the regional controller. If approved, a coupling policy is issued, defining key parameters such as timing, leader-follower roles, and communication mode. The edge controller then initiates the coupling process by establishing a V2V/V2I communication link, synchronizing positions, and enabling real-time sharing of movement intentions. This allows the Pods to operate in a coordinated formation while maintaining safety through continuous radar/GNSS data exchange. Once the shared route ends, decoupling is automatically triggered, and the event is logged for later evaluation. This process enables coordinated multi-Pod formation without continuous central supervision as the edge controller manages the full coupling lifecycle.

**Adaptive Interface Switching.** This mechanism addresses scenarios where a Pod experiences communication degradation or sudden environmental changes, such as link failure or road incidents. The edge controller periodically monitors each Pod's link quality and, upon detecting degradation, switches to an alternative interface to maintain connectivity. In case of emergencies or complete link failures, the Pod node is also able to autonomously evaluate available interfaces and re-establish a minimal communication link to report its status. Once conditions stabilize, the controller verifies system state and either restores the original configuration or retains the new interface if it offers better performance. This mechanism ensures resilient connectivity and localized emergency handling, even when central coordination is temporarily disrupted.

## 5 Conclusions and Future Perspectives

This paper presented a conceptual architecture for the Pods4Rail system, separating scheduling, collaboration, and transport layers to support modular and adaptive communication. Moreover, it introduced a distributed SDN management framework by combining regional coordination with edge autonomy. To demonstrate how distributed intelligence can ensure reliable, low-latency communication in multimodal transportation systems, two control mechanisms were developed: virtual coupling of Pods and adaptive RAT switching. Future work will extend this foundation in several directions. At the system level, simulation and prototyping will validate performance under large-scale, dynamic conditions. At the node level, we plan to refine SDN control granularity to better manage interface switching in heterogeneous networks. Additionally, future work will address the security implications of distributed control, including mutual authentication of edge controllers and integrity verification of flow rule distribution.

**Acknowledgements**

The work presented is funded by ADEME project FLEXMOVE and the European Union under Pods4Rail project, GA 101121853. Views and opinions expressed are however those of the author(s) only and do not necessarily reflect those of the European Union or Europe's Rail Joint Undertaking. Neither the European Union nor the granting authority can be held responsible for them.